\documentclass[aps,preprint,amsmath,amssymb,citeautoscript,showpacs]{revtex4}

\usepackage{graphicx,times}
\usepackage{amsmath,amssymb,amsbsy,amsfonts}
\usepackage{array}
\usepackage{bm}
\usepackage{graphics}
\usepackage{mathrsfs}
\usepackage{xcolor}
\usepackage{cancel}
\usepackage[normalem]{ulem}

\newcommand{\bfS}{{\mathbf{S}}}

\newcommand{\ua}{{\uparrow}}
\newcommand{\da}{{\downarrow}}

\begin{document}

\title{Solution to the sign problem in a frustrated quantum impurity model}
\author{Connor T. Hann$^a$, Emilie Huffman$^a$ and Shailesh Chandrasekharan$^{a,b}$}
\affiliation{$^a$Department of Physics, Box 90305, Duke University,
Durham, NC 27708, USA \\
$^b$Center for High Energy Physics, Indian Institue of Science, Bangalore, 560 012, India}

\begin{abstract}
In this work we solve the sign problem of a frustrated quantum impurity model consisting of three quantum spin-half chains interacting through an anti-ferromagnetic Heisenberg interaction at one end. We first map the model into a repulsive Hubbard model of spin-half fermions hopping on three independent one dimensional chains that interact through a triangular hopping at one end. We then convert the fermion model into an inhomogeneous one dimensional model and express the partition function as a weighted sum over fermion worldline configurations. By imposing a pairing of fermion worldlines in half the space we show that all negative weight configurations can be eliminated. This pairing naturally leads to the original frustrated quantum spin model at half filling and thus solves its sign problem.
\end{abstract}


\maketitle

\section{Introduction}
\label{sec1}

Understanding quantum many body physics, especially in the regime where the degrees of freedom are strongly correlated, is one of the outstanding areas of research in both condensed matter and nuclear physics today. Many physical phenomena ranging from the properties of nuclei \cite{Bedaque:2002mn}, phases of dense nuclear matter \cite{Schmitt:2010pn}, properties of high $T_c$ materials \cite{And97}, heavy fermion systems \cite{HFSys}, topological superconductors and insulators \cite{RevModPhys.83.1057}, etc., contain strongly correlated regimes of interest. While approximate methods can help uncover exotic features that can emerge in such systems, reliable quantitative predictions usually require numerical approaches such as the quantum Monte Carlo method \cite{Fehske08,Lee:2008fa,Drut:2012md,RevModPhys.87.1067}. Unfortunately these methods suffer from sign problems in many interesting cases, and their solutions are exciting research directions in the field of computational quantum many body theory today \cite{Gattringer:2016kco}.

The challenge is to rewrite the quantum problem as a classical statistical mechanics problem with positive Boltzmann weights that are computable in polynomial time. While the Feynman path intergal is one way to proceed, the presence of fermions and/or frustration means there is no guarantee that the goal can be achieved. Although a generic solution that solves all sign problems most likely does not exist \cite{Tro05}, solutions to many specific sign problems have been discovered recently \cite{PhysRevLett.83.3116,Alford:2001ug,PhysRevLett.99.250201,PhysRevD.85.091502,Bloch:2011jx,PhysRevB.77.125101,Bloch:2015iha,PhysRevLett.115.266802,Chandrasekharan:2012fk,Huf14,Li15,Wan15a,PhysRevLett.116.250601,Li16,Chandrasekharan:2013rpa,Huffman:2016nrx,Alet16alm,PhysRevB.93.054408}. It would be nice to establish general criteria for the solvability of sign problems.

Discovering an appropriate basis to formulate the quantum problem is an important step in the solution to the sign problem in a given system. For example, the sign problem that exists in a class of frustrated quantum spin systems when formulated in the local spin-half basis can be eliminated by going to a local spin-one basis \cite{Alet16alm, PhysRevB.93.054408}. It would be exciting if a systematic approach could be developed to construct such a basis for each problem of interest. Every new solution expands the class of solvable problems and thus takes us a step closer towards this goal. In this work we discover a solution to a simple frustrated model involving a single triangular anti-ferromagnetic interaction. In order to make the problem non-trivial, each spin in the triangle is coupled to its own bath of spins in the form of a one-dimensional chain. Our model was considered earlier as a toy model to explore if a basis change could help alleviate the sign problem \cite{PhysRevB.92.195126}. It was shown that even a small change in the basis can have a significant effect on alleviating the sign problem. In this work we explore if the sign problem can in fact be completely eliminated in certain cases. Although our model is geometrically different from those considered in \cite{Alet16alm, PhysRevB.93.054408}, our final solution is similar and emerges when the system is formulated in a local spin-one basis on half the system. Interestingly, we can go a step further and show that the solution is based on fermion pairing in a related fermion model. 

Our paper is organized as follows. In section \ref{sec2} we explain the details of our model and map it into a fermion model which plays an important role in uncovering our solution. In section \ref{sec3} we transform the model into an inhomogeneous one dimensional model by identifying new fermion degrees of freedom on half the lattice. We then expand the partition function using fermion worldlines and identify the origin of the sign problem. In section \ref{sec4} we show that the sign problem is absent in a model that contains only paired fermion worldlines. Using this insight, in section \ref{sec5} we define a new local basis for the original spin model that is free of sign problems. Section \ref{sec6} contains our conclusions.

\begin{figure}[b]
\includegraphics[width=0.8\textwidth]{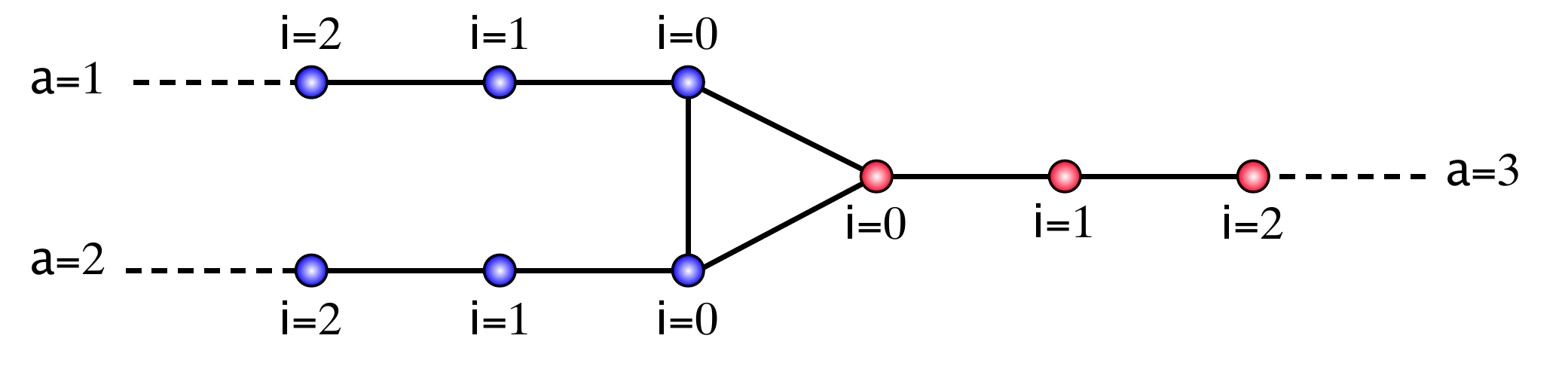}
\caption{\label{fig1} The lattice structure of the frustrated quantum impurity model. }
\end{figure}

\section{Frustrated Quantum Impurity Model}
\label{sec2}

In this work we consider a model consisting of three quantum spin chains constructed with spin half operators $\bfS_{a,i}$, where $a=1,2,3$ labels the three chains and $i=0,...,N$ labels the sites in each chain. Frustration is introduced through an anti-ferromagnetic interaction among the three quantum spins at the $i=0$ site  (see Fig.~\ref{fig1}). The Hamiltonian of the system is given by
\begin{equation}
H \ =\  \lambda_s \sum_{a=1,2,3} \sum_{i=0}^{N-1} \bfS_{a,i} \cdot \bfS_{a,i+1}\ + 
\alpha_s \big(\bfS_{1,0}\cdot \bfS_{2,0} + (\bfS_{1,0}+\bfS_{2,0})\cdot \bfS_{3,0}\big)
\label{smodel}
\end{equation}
where $\lambda_s$ and $\alpha_s$ are two independent couplings. The spin-half operators satisfy the usual commutation relations $[S^x_{a,i},S^y_{b,j}] = i S^z_{a,i}\delta_{ab} \delta_{ij}$. Our goal is to write the partition function
\begin{equation}
Z \ =\ \mathrm{Tr}\Big(\ \mathrm{e}^{-\beta H}\ \Big)
\end{equation}
 as a sum over configurations with positive weights, such that each weight is computable in polynomial time as the system size (i.e., $N$) and $\beta$ grow. Construction of such an expansion is referred to as a solution to the sign problem for the quantum system described by $H$. For general quantum systems the existence of such an expansion is not guaranteed, especially in the presence of frustrating interactions. The idea is to explore solutions to sign problems in simple models as a step towards finding similar solutions in more complex models. Our model contains only a simple local frustration due to the triangular anti-ferromagnetic coupling $\alpha_s \neq 0$, but it is sufficient to introduce sign problems with conventional methods. In contrast, in this work we wish to solve the sign problem completely. Our model does not naturally fall in the class of frustrated models solved recently in~\cite{Alet16alm,PhysRevB.93.054408}, but we will show that the final solution, which relies on the mapping to an inhomogeneous one dimensional system, is similar. 

In our approach, we first map the spin model into a fermion model whose creation and annihilation operators of are denoted as $c^\dagger_{\sigma,a,i}, c_{\sigma,a,i}$ where, as before, the index $a$ denotes the chain, the index $i$ denotes the lattice site on each chain, and $\sigma=\uparrow,\downarrow$ are the spin degrees of freedom. The site $i=0$ is special and viewed as an impurity site that couples the three different lattices.  At this site fermions can hop between the three different lattices. It is easy to show that the Hamiltonian
\begin{equation}
H= - \lambda \sum_{a,\sigma}\ \sum_{i=0}^{N-1} 
(c^\dagger_{\sigma,a,i} c_{\sigma,a,i+1} +c^\dagger_{\sigma,a,i+1} c_{\sigma,a,i})
- \alpha \sum_{\sigma,a,b}\ c^\dagger_{\sigma,a,0} M_{ab}\ c_{\sigma,b,0} - U \sum_{a,i} Q_{a,i}.
\label{fmodel}
\end{equation}
where
\begin{equation}
M = \left(\begin{array}{ccc} 0 & 1 & 1 \cr 1 & 0 & 1 \cr 1 & 1 & 0 \end{array}\right),
\end{equation}
and
\begin{equation}
Q_{a,i} \ =\ 
\big(c^\dagger_{\uparrow,a,i}c_{\uparrow,a,i}-c^\dagger_{\downarrow,a,i}c_{\downarrow,a,i}\big)^2,
\label{qdef}
\end{equation}
will reproduce the physics of (\ref{smodel}) with $\lambda_s = 2 \lambda^2/U$ and $\alpha_s = 2 \alpha^2/U$ in the limit of $U \rightarrow \infty$. In this limit the system is forced to contain a single fermion degree of freedom at every lattice site. Our initial hope was that if we could solve the sign problem in this fermionic model, it would be more general and exciting. While we have not yet found a full solution, the sign problem in the fermion model is eliminated if we impose additional pairing in the fermion worldline configurations. In the large $U$ limit this pairing leads to a class of quantum spin models in which our original impurity model is included. Using insight from this pairing solution we can construct a basis in the original spin model that solves its sign problem.

\begin{figure}[t]
\includegraphics[width=0.5\textwidth]{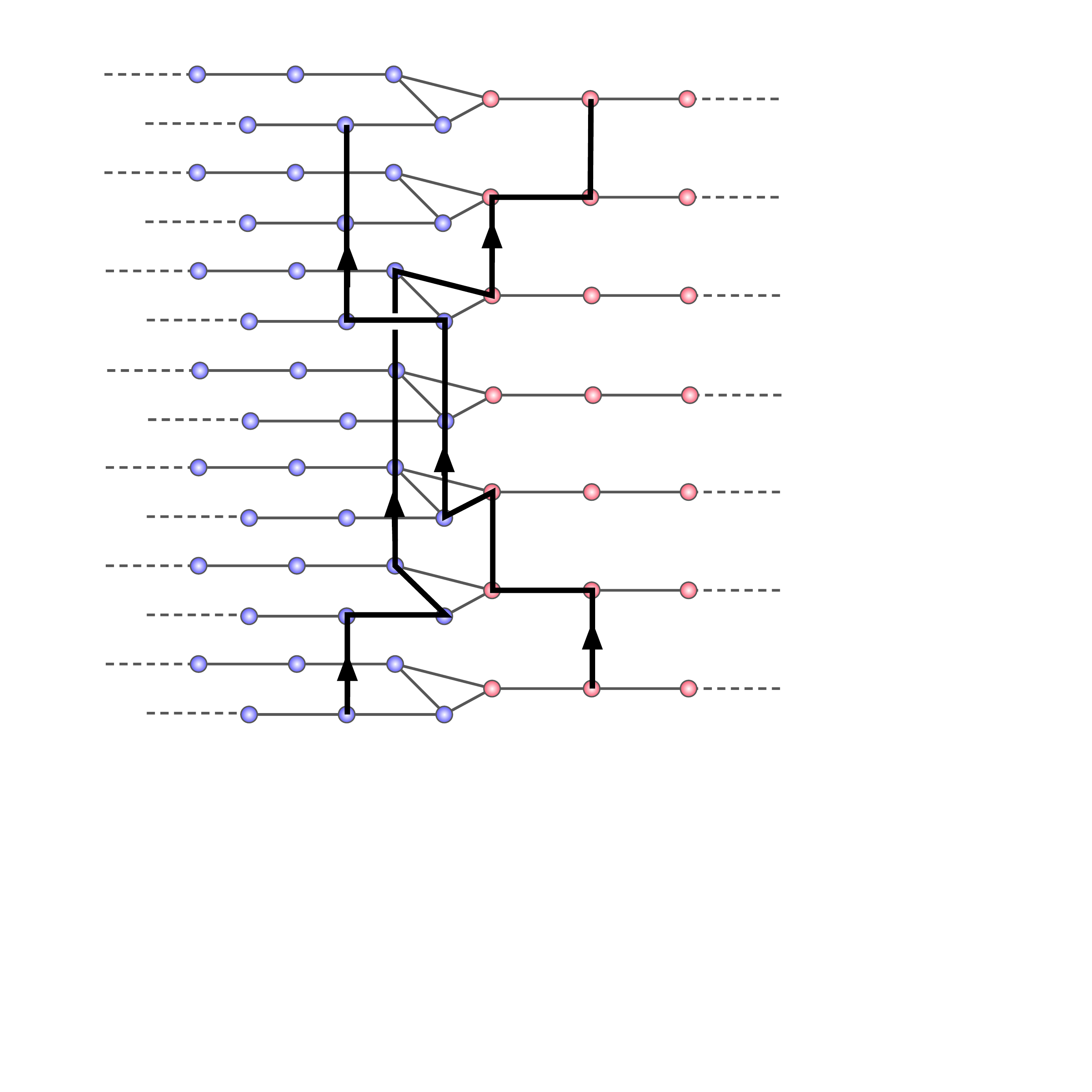}
\caption{\label{fig2} A fermion worldline configuration showing permutation of two fermions.}
\end{figure}

\section{Transforming to $S$ and $A$ type fermions}
\label{sec3}

Solutions to sign problems in Fermi systems are difficult since sign problems can arise through fermion permutations. For example, model (\ref{fmodel}) contains a sign problem in the fermion occupation number basis even when $U=0$. Indeed, worldlines of fermions can permute with each other by hopping around the triangle as illustrated in the Fig. \ref{fig2}. Hence, solutions to fermion sign problems usually require resummations of world line configurations which lead to fermion determinants. Such solutions to sign problems are available for half filled systems with repulsive interactions such as ours, but only on bi-partite lattices. Unfortunately, this conventional solution no longer works for model (\ref{fmodel}) due to the triangular hopping term proportional to $\alpha$. Thus, even our simple system offers an opportunity to explore solutions to sign problems that go beyond conventional methods.

Open one dimensional fermion systems with nearest neighbor hopping have no sign problems in the occupation number basis since fermion worldlines of the same type cannot cross each other due to the Pauli principle. In this work we wish to explore if this idea can be used to solve the sign problem in our model. For this purpose let us define a new set of fermion creation operators using the following unitary transformation:
\begin{equation}
\left(\begin{array}{c} c_{\sigma,1,i}^\dagger \cr c_{\sigma,2,i}^\dagger \cr c^\dagger_{\sigma,3,i}\end{array}\right)
\ =\ 
\left(\begin{array}{ccc} 1/\sqrt{2} & 1/\sqrt{2} &  0 \cr -1/\sqrt{2} & 1/\sqrt{2} & 0 \cr 0 & 0 & 1 \end{array}\right)\ 
\left(\begin{array}{c} d_{\sigma,1,i}^\dagger \cr d_{\sigma,2,i}^\dagger\cr d^\dagger_{\sigma,3,i} \end{array}\right).
\end{equation}
Such a transformation has already been demonstrated to minimize the severity of the sign problem in frustrated impurity models~\cite{PhysRevB.92.195126}. The annihilation operators are defined using the Hermitian conjugate of the above expression. In terms of $d^\dagger_{\sigma,a,i}$ and $d_{\sigma,a,i}$ the free part of (\ref{fmodel}) can be written as
\begin{equation}
H_0 = -\lambda \sum_{a,\sigma}\ \sum_{i=0}^{N-1} 
(d^\dagger_{\sigma,a,i} d_{\sigma,a,i+1} +d^\dagger_{\sigma,a,i+1} d_{\sigma,a,i})
- \alpha \sum_{\sigma,a,b}\ d^\dagger_{\sigma,a,0} K_{ab}\ d_{\sigma,b,0}
\label{1dch}
\end{equation}
where
\begin{equation}
K = \left(\begin{array}{ccc} -1 & 0 & 0 \cr 0 & 1 & \sqrt{2} \cr 0 & \sqrt{2} & 0 \end{array}\right).
\end{equation}
Thus, the above transformation helps convert the free part of the three coupled chains into two disconnected open chains, one involving nearest neighbor hops of the $S$-type fermions (created by $d^\dagger_{\sigma,2,i}$ and $d^\dagger_{\sigma,3,i}$) and the other involving nearest neighbor hops of the $A$-type particle (created by $d^\dagger_{\sigma,1,i}$). We can view the transformed basis as though we have combined the Hilbert spaces on each site in the $a=1$ and $a=2$ chains into a single {\em composite} site, while each site in the $a=3$ chain remains as a {\em fundamental} site. Thus, the system looks one dimensional (see Fig.~\ref{fig2}) but inhomogeneous. The composite sites (shown as squares) contain a sixteen dimensional Hilbert space as compared to fundamental sites (shown as circles), which have a four dimensional Hilbert space.
\begin{figure}[h]
\includegraphics[width=0.8\textwidth]{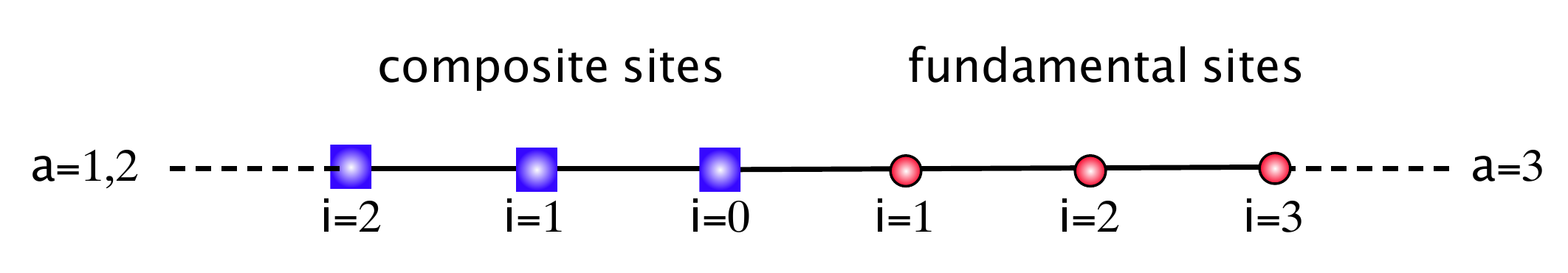}
\caption{\label{fig3} Combining the Hilbert spaces at the site $i$ in the a=1 and a=2 chains into a single composite site one obtains a system that looks one-dimensional but inhomogeneous. The lattice sites have been relabeled as explained in the text.}
\end{figure}
The $S$-type particles move on the full one dimensional chain, while the $A$-type particles are restricted to the half chain consisting of composite sites. To reflect the one dimensional nature of the new lattice, we relabel the lattice sites to a one dimensional chain $-N \leq i \leq N+1$, where $i \leq 0$ label composite sites and $ i \geq 1$ label fundamental sites. We also  define new $A$-type and $S$-type fermion operators with the one-dimensional site index as follows: 
\begin{equation}
b^\dagger_{\sigma,A,-i} \ =\  d^\dagger_{\sigma,1,i},\quad
b^\dagger_{\sigma,S,-i} \ = \ d^\dagger_{\sigma,2,i}, \quad
b^\dagger_{\sigma,S,i+1} \ = \ d^\dagger_{\sigma,3,i} 
\end{equation}
where $i=0,1,2...,N$. 

Using the new fermion creation and annihilation operators (\ref{fmodel}) can be rewritten as
\begin{equation}
H = H_0^A + H_0^S +  H_0^{LR}+ H_\mu+H_{\rm int}
\end{equation}
where
\begin{subequations}
\begin{eqnarray}
H_0^A &=& -\lambda \sum_{\sigma,i=-N}^{-1} \ \big(b^\dagger_{\sigma,A,i}b_{\sigma,A,i+1} + b^\dagger_{\sigma,A,i+1}b_{\sigma,A,i}\big), \\
H_0^S &=& -\lambda \Big\{\sum_{\sigma,i=-N}^{-1} \ + \sum_{\sigma,i=1}^{N} \Big\}\ 
\big(b^\dagger_{\sigma,S,i}b_{\sigma,S,i+1} + b^\dagger_{\sigma,S,i+1}b_{\sigma,S,i}\big), \\
H_0^{LR} &=& - \sqrt{2}\alpha\big(b^\dagger_{\sigma,S,0}b_{\sigma,S,1} + b^\dagger_{\sigma,S,1}b_{\sigma,S,0}\big),
\\
H_\mu &=& -\alpha \sum_\sigma\ (b^\dagger_{\sigma,S,0}b_{\sigma,S,0}  - b^\dagger_{\sigma,A,0}b_{\sigma,A,0}),
\\
H_{\rm int} &=& - U\sum_{n,i=-N}^{i=N+1}  Q^{n}_i. \label{Hint}
\end{eqnarray}
\label{model1}
\end{subequations}
Note that $H^{LR}$ represents hopping between the composites on the left and fundamental sites on the right and only involves $S$-type particles. The interaction operators $Q^{n}_i$ can be constructed using (\ref{qdef}) and will differ between fundamental and composite sites due to the difference in the Hilbert space. On the composite sites we will write them as a sum over five different operators labeled $Q^{1}_i,Q^{2,+}_i,Q^{2,-}_i,Q^{2,0}_i,Q^{3}_i$, while on the fundamental sites we will need only one such operator $Q^{1}_i$. We will construct these individual operators explicitly after choosing a basis to expand the partition function.

We expand the partition function of the model in the fermion occupation number basis defined by ordering the creation of fermions in the Fock vacuum. We first create fermions on the right most lattice site (i,e...$i=N+1$) and then move to the left. For fundamental sites $(i\geq 1)$, we define the four dimensional basis states as
\begin{equation}
|n_1n_2\rangle = (b^\dagger_{\ua,S,i})^{n_1}(-b^\dagger_{\da,S,i})^{n_2}|0\rangle,
\label{basisf}
\end{equation}
where, for later convenience, we have introduced an extra negative sign when particles of spin down are created. Similarly, for composite sites $(i\leq 0)$ we choose the occupation number basis as
\begin{equation}
|n_1n_2n_3n_4\rangle = (b^\dagger_{\ua,A,i})^{n_1}(b^\dagger_{\ua,S,i})^{n_2} (b^\dagger_{\da,A,i})^{n_3}(-b^\dagger_{\da,S,i})^{n_4}|0\rangle,
\label{basisc}
\end{equation}
i.e, create up spins after down spins and if there are two particles with the same spin create $S$ type particles before $A$ type ones. Again note the extra negative sign in front of the $ b^\dagger_{\da,S,i}$ as before.

\begin{table}[h]
\begin{tabular}{|l||l|}
\hline
$H_\mu\|0001\rangle = -\alpha|0001\rangle$ & $H_\mu\|0010\rangle = \alpha|0010\rangle$ \\
$H_\mu\|0100\rangle = -\alpha|0100\rangle$ & $H_\mu\|1000\rangle = \alpha|1000\rangle$ \\
$H_\mu|0111\rangle = -\alpha|0111\rangle$ & $H_\mu|1011\rangle = \alpha|1011\rangle$ \\
$H_\mu|1101\rangle = -\alpha|1101\rangle$ & $H_\mu|1110\rangle = \alpha|1110\rangle$ \\
$H_\mu|1010\rangle =  2\alpha|1010\rangle$ & $H_\mu|0101\rangle =  -2\alpha|0101\rangle$ \\
\hline
\end{tabular}
\caption{\label{tab1} The operator $H_\mu$ is diagonal in the fermion occupation number basis. The above table gives the non-zero eigenvectors and the corresponding eigenvalues.}
\end{table}

\begin{table}[h]
\begin{tabular}{|l||l|}
\hline
$(Q_{1,i}+Q_{2,i})|0000\rangle = 0$ & $(Q_{1,i}+Q_{2,i})|1111\rangle = 0$ \\
$(Q_{1,i}+Q_{2,i})|0001\rangle = |0001\rangle$ & $(Q_{1,i}+Q_{2,i})|0010\rangle = |0010\rangle$ \\
$(Q_{1,i}+Q_{2,i})|0100\rangle = |0100\rangle$ & $(Q_{1,i}+Q_{2,i})|1000\rangle = |1000\rangle$ \\
$(Q_{1,i}+Q_{2,i})|0111\rangle = |0111\rangle$ & $(Q_{1,i}+Q_{2,i})|1011\rangle = |1011\rangle$ \\
$(Q_{1,i}+Q_{2,i})|1101\rangle = |1101\rangle$ & $(Q_{1,i}+Q_{2,i})|1110\rangle = |1110\rangle$ \\
$(Q_{1,i}+Q_{2,i})|1100\rangle = 2 |1100\rangle$ & $(Q_{1,i}+Q_{2,i})|0011\rangle = 2|0011\rangle$ \\
$(Q_{1,i}+Q_{2,i})|1010\rangle =  (|1010\rangle+|0101\rangle)$ & $(Q_{1,i}+Q_{2,i})|0101\rangle =  (|0101\rangle+|1010\rangle)$ \\
$(Q_{1,i}+Q_{2,i})|1001\rangle = (|1001\rangle+|0110\rangle)$ & $(Q_{1,i}+Q_{2,i})|0110\rangle = (|0110\rangle+|1001\rangle)$ \\
\hline
\end{tabular}
\caption{\label{tab2} Action of the interaction term $(Q_{1,i}+Q_{2,i})$ on the sixteen dimensional Hilbert space on a composite site. Most of the Hilbert space states are eigenstates of the interaction operator. Only two sets of doublets mix within each doublet.}
\end{table}

In the above basis we can compute the matrix elements of various terms that appear in the Hamiltonian (\ref{model1}). First note that $H_\mu$ is diagonal in the occupation number basis, where the non-zero diagonal elements are given in Table.~\ref{tab1}. Next we consider the interaction term and write it as $H_{\rm int} = H^C_{\rm int} + H^F_ {\rm int}$, where $H^C_{\rm int} \ =\ -U\ \sum_i Q_{1,i}+Q_{2,i}$ and $H^F_{\rm int} \ =\ -U\ \sum_i Q_{3,i}$ represent the interactions in the composite and fundamental parts of the chain. For the fundamental chain $Q_{3,i}$ is diagonal in our chosen basis:
\begin{equation}
Q_{3,i}|00\rangle = Q_{3,i}|11\rangle = 0, \ \  Q_{3,i}|01\rangle = |01\rangle,\ \ 
Q_{3,i}|10\rangle = |10\rangle.
\end{equation}
If we define $Q_i^1$, which appears in (\ref{Hint}), on the fundamental sites to be a projector on the one particle space, such that $Q_i^1|n_1 n_2\rangle = \delta_{n_1+n_2,1}|n_1 n_2\rangle$, note that $Q_i^1 \equiv Q_{3,i}$. For the composite sites, the action of the interaction term $(Q_{1,i}+Q_{2,i})$ on the sixteen dimensional basis states is given table \ref{tab2}. We define $Q_i^n |n_1n_2n_3n_4\rangle =  \delta_{n,n_1+n_2+n_3+n_4}|n_1n_2n_3n_4\rangle$ for $n=1,3$. For $n=2$ we define three different operators with the property that 
\begin{equation}
Q_i^{2,+} |1100\rangle = 2|1100\rangle,\quad  Q_i^{2,-} |0011\rangle = 2|0011\rangle,
\end{equation}
\begin{subequations}
\vspace{-1.15cm}
\begin{eqnarray}
Q_i^{2,0}|1010\rangle =  (|1010\rangle+|0101\rangle), && 
Q_i^{2,0}|0101\rangle =  (|1010\rangle+|0101\rangle) \\
Q_i^{2,0}|1001\rangle =  (|1001\rangle+|0110\rangle), && 
Q_i^{2,0}|0110\rangle =  (|1001\rangle+|0110\rangle) 
\end{eqnarray}
\label{q20}
\end{subequations}
Using these definitions we see that $(Q_{1,i}+Q_{2,i}) = Q_i^1 + Q_i^{2,+} + Q_i^{2,-} +Q_i^{2,0} + Q_i^3$. Finally we turn to the action of each of the nearest neighbor hopping terms contained in  $H_0^A$, $H_0^S$, $H_0^{LR}$. We denote these hops generically as $H^{a,\sigma}_{ij} = -t_{ij} b^\dagger_{\sigma,a,i}b_{\sigma,a,j}$, which hops an $a=A,S$ type fermion of spin $\sigma$ from site $j$ to $i$. Here $t_{ij}$ is $\lambda$ or $\sqrt{2}\alpha$ depending on the bond $ij$.  

\begin{figure}[h]
\includegraphics[width=0.7\textwidth]{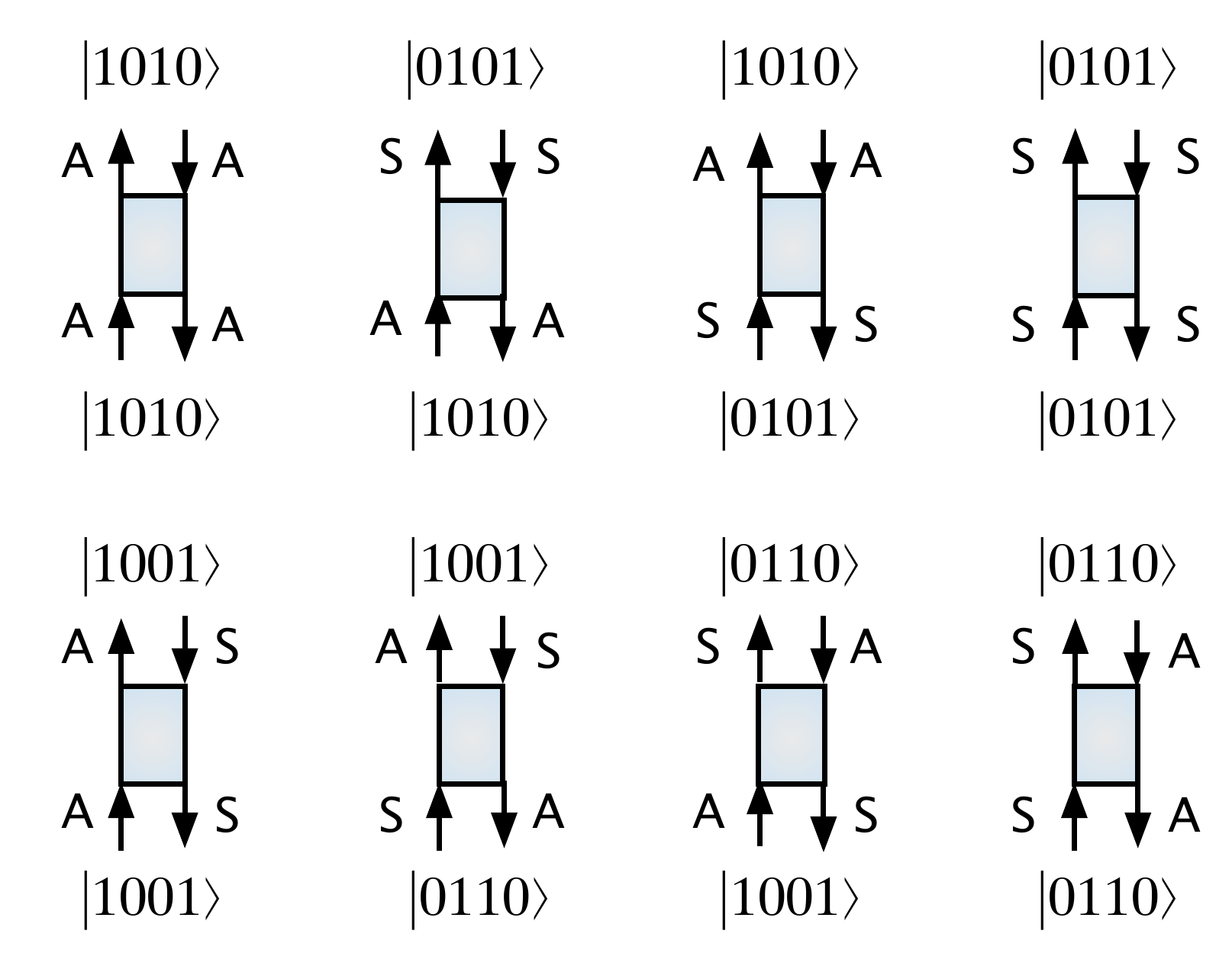}
\caption{\label{fig4} The eight non-zero matrix elements of the operator $Q_i^{2,0}$, which can induce a transition between an $S$ type to an $A$ type fermion of the same spin. These matrix elements are all positive as can be seen from Eq.~(\ref{q20}).}
\end{figure}

Having computed all the matrix elements we can expand the partition function in the CT-INT representation given by \cite{Rub05},
\begin{equation}
Z = \sum_k \int_T dt_1 dt_2...dt_k (-1)^k \mathrm{Tr}\Bigg(\mathrm{e}^{-(\beta-t_1)H_\mu}H_1\mathrm{e}^{-(t_1-t_2)H_\mu}H_2...H_3...H_k\mathrm{e}^{-t_kH_\mu}\Bigg)
\label{ctint}
\end{equation}
where the integral is over the time ordered domain $T \equiv t_1 \geq t_2 \geq,...\geq t_k$, and $H_p$ stands for one of the many possible local terms in the Hamiltonian like the fermion hop $H^{a,\sigma}_{ij}$ or $-U Q_i^{n}$. We then compute the trace by introducing a sum over fermionic occupation basis between various $H_p$ insertions. While the hopping terms move the fermions around, all other terms except $Q_i^{2,0}$ are diagonal terms and do not change the state in our chosen basis of fermion occupation numbers. The action of $Q_i^{2,0}$ on sites $i\leq 0$ is shown pictorially in Fig.~\ref{fig4}. We see that there are eight non-zero matrix elements, four of them diagonal while the other four flip an $A$-type particle into a $S$-type particle and vice versa without changing the spin. Hence, while we can follow the world line of a particular spin, S-type and A-type fermions mix among themselves due to $Q_i^{2,0}$. Worldlines of a specific spin are still well defined, and pictorially the trace becomes a sum over these worldline configurations $[\ell]$:
\begin{equation}
Z = \sum_k \int_T dt_1 dt_2...dt_k \sum_{[\ell]} \mathrm{Sign}([\ell]) W([\ell])
\end{equation}
where 
\begin{equation}
W([\ell]) = \lambda^{k_1} (\sqrt{2}\alpha)^{k_2} (2U)^{k_3} (U^{k_4}) \Omega_\mu 
\end{equation}
where $k_1$ stands for the number of fermion hops coming from $H_0^A$ or $H_0^S$ terms, $k_2$ stands for the number of fermion hops coming from $H_0^{LR}$, $k_3$ stands for the number of insertions of $Q_i^{2,+}$ and $Q_i^{2,-}$ and $k_4$ stands for the number of insertions of the other $Q_i^{n}$. The factor $\Omega_\mu$ is a postitive number that depends on the $H_\mu$ term. Whenever there is a $S$ type particle that exists for a time $\tau$ it contributes a factor $\exp(\alpha \tau)$ to it, on the other hand an $A$ type particle contributes $\exp(-\alpha\tau)$ to it.

\begin{figure}[h]
\includegraphics[width=0.7\textwidth]{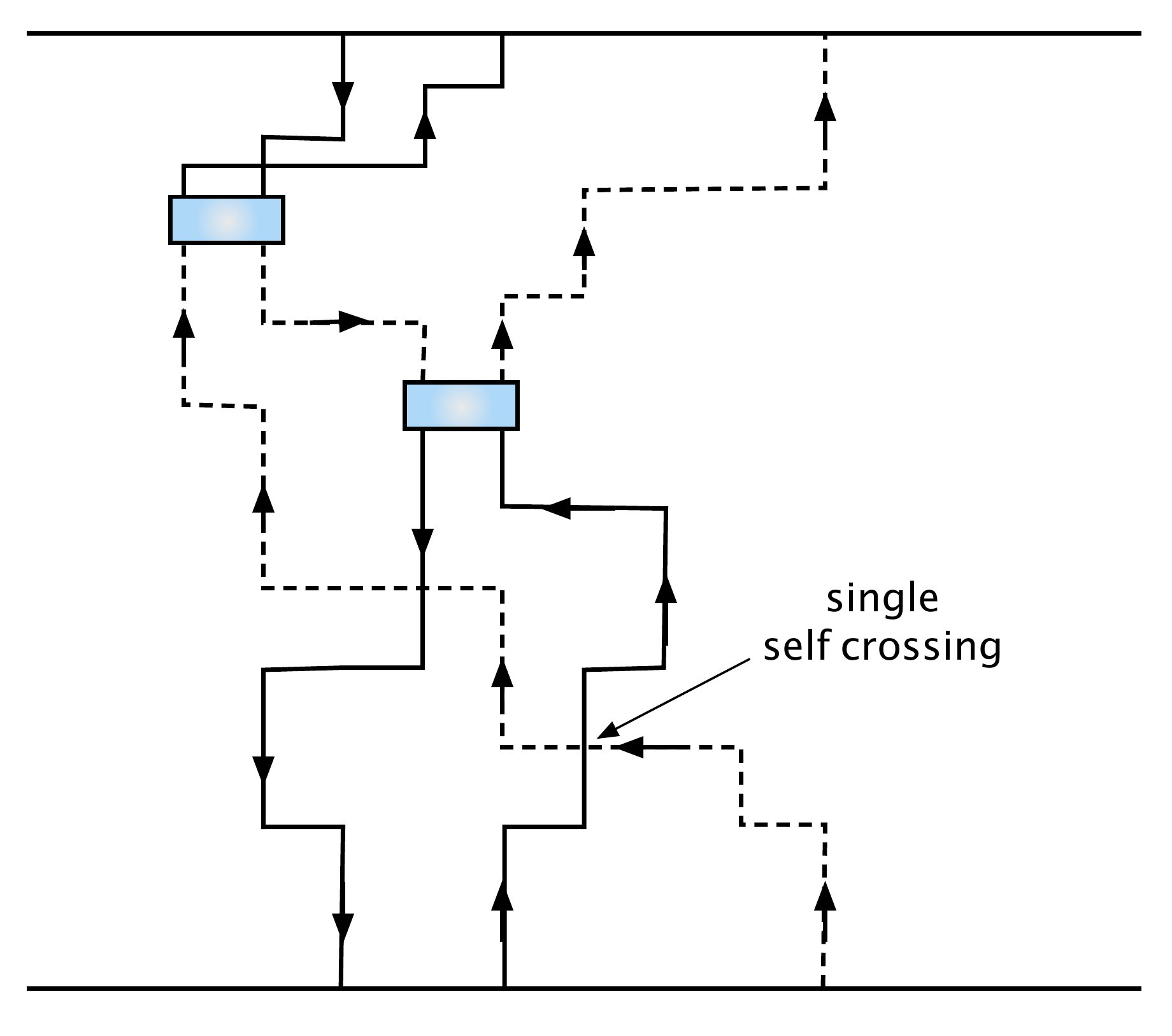}
\caption{\label{fig5} Example of a negative weight fermion worldline configuration. The configuration contains one spin-up ($\uparrow$) loop that self intersects once and one spin-down ($\downarrow$) loop that does not self intersect. $S$ type and $A$ type particles are indicated by dashed and solid lines respectively, while the blue boxes are instances of the matrix elements shown in Fig.~\ref{fig4}.}
\end{figure}

Fermion worldline configurations $[\ell]$ are a set of closed loops that can wrap over the temporal boundary. The sign of a configuration, $\mathrm{Sign}([\ell])$, comes from the number of crossings of fermion world lines. In one dimension any two different loops will cross an even number of times. Hence, negative factors from crossings of different loops always give a positive sign. This means the sign of a configuration is determined by the number of self crossings of each loop, which we can denote by $N_{\rm self}$. Then
\begin{equation}
\mathrm{Sign}([\ell]) = (-1)^{N_{\rm self}}.
\end{equation}
It is easy to build configurations with negative weight as shown in Fig.~\ref{fig5}. Thus, although in the absence of interactions there is no sign problem, interactions do introduce configurations with negative signs. It would be interesting if this sign problem can be solved using methods like the meron cluster \cite{PhysRevLett.83.3116} or the fermion bag approach. While we have made much progress in this direction, the complete solution is still missing. Hence, we postpone this discussion to a later publication. Instead, in this work we try to identify configurations that arise in the limit $U \rightarrow \infty$ and show that these are guaranteed to be positive and thus solve the sign problem in the original quantum spin model. 

\section{Paired Fermion Model}
\label{sec4}

Fermion sign problems can often be solved by pairing fermion worldlines or when fermions are one dimensional. A combination of both these features are required to solve the sign problem in (\ref{smodel}). In the limit of $U\rightarrow \infty$, the $H_{\rm int}$ term in (\ref{fmodel}) dominates and forces every composite site to have exactly two fermions and every fundamental site to have a single particle. Further, out of the six possible two particle states involving paired fermions, the operator, $(Q_{1,i}+Q_{2,i})$ projects into four states $t^+$, $t^-$, $t^0$ and $s$ defined by
\begin{equation}
|t^+\rangle = |1100\rangle,\ \ 
|t^-\rangle = |0011\rangle,\ \ 
|t^0\rangle = \frac{1}{\sqrt{2}}\big(|0110\rangle+|1001\rangle\big),\ \ 
|s\rangle = \frac{1}{\sqrt{2}}\big(|1010\rangle+|0101\rangle\big),\ \ 
\label{4states}
\end{equation}
as can be seen from table \ref{tab2}. We will show that restricting the Hilbert space on the composite sites to these paired states eliminates all negative weight configurations due to the one dimensional nature of the full problem. Interestingly, the weights remain positive even even if we allow the composite sites to be empty without any fermions. Thus, the fermion model can be modified by adding the additional interaction term
\begin{equation}
H'_{\rm int} = H_{\rm int} - U\sum_{i} Q_i^{0},
\end{equation}
where $Q_i^{0} |n_1,n_2\rangle = \delta_{n_1+n_2,0} |n_1,n_2\rangle$ on fundamental sites and $Q_i^{0} |n_1,n_2,n_3,n_4\rangle = 2\delta_{n_1+n_2+n_3+n_4,0} |n_1,n_2,n_3,n_4\rangle$ on composite sites. With the additional term, the $U\rightarrow \infty$ limit defines a paired fermion model which is similar to the original frustrated quantum spin model (1) but now allows composite sites and fundamental sites to be empty. In this model fermion worldline configurations only contain paired fermions on the composite sites and unpaired fermions on the fundamental sites, along with empty sites. In fact we will be more general and define the model through these worldline configurations and their weights. Such models are difficult to write down explicitly in terms of a Hamiltonian, so we do not attempt it here. We now focus on this more general model, bearing in mind that the original quantum spin model is what results when all empty sites are eliminated.

\begin{figure}[h]
\includegraphics[width=0.7\textwidth]{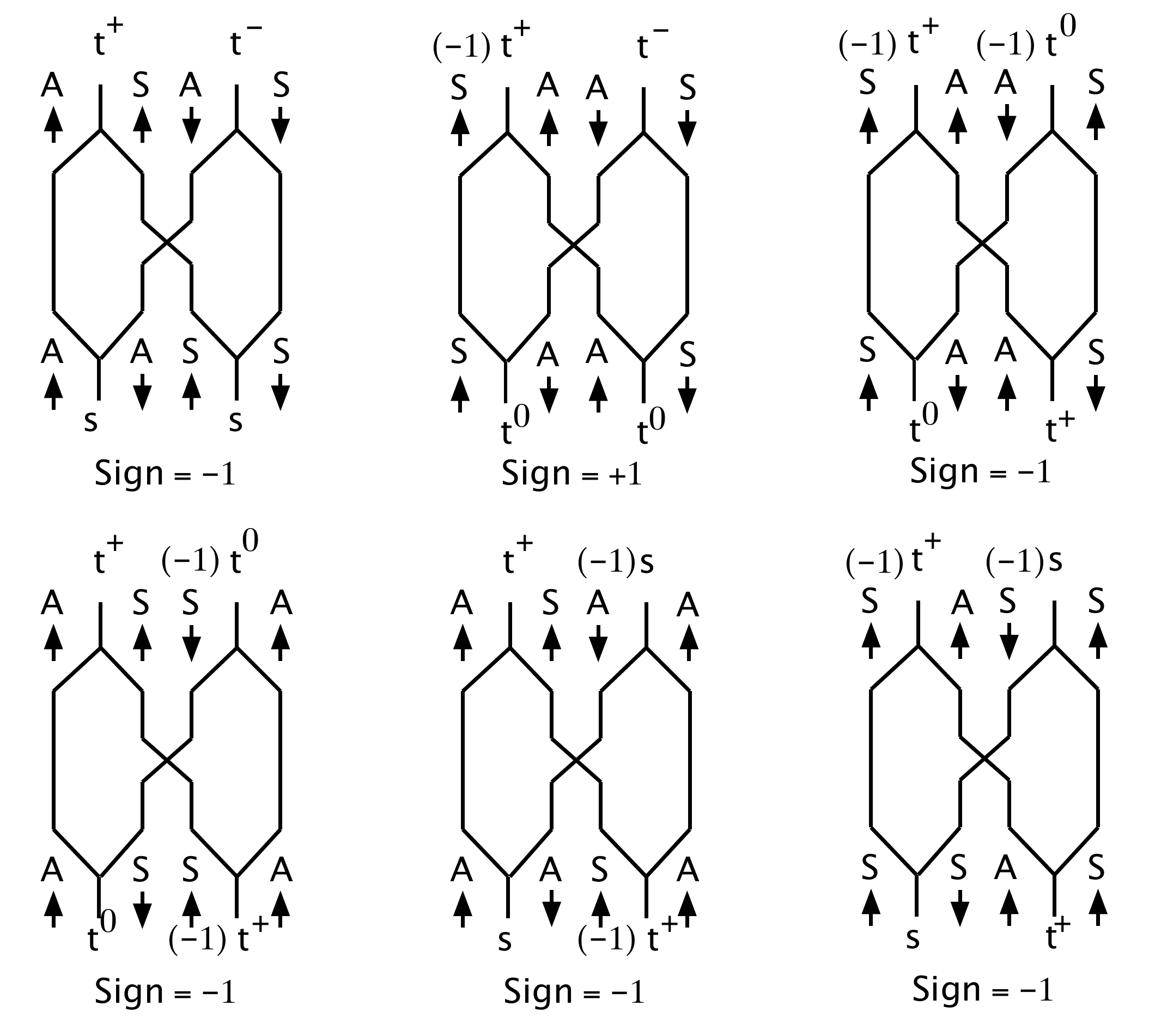}
\caption{\label{fig6} Examples of fermion exchange diagrams and the sign change associated the processes listed in Table \ref{tab3}. The sign of each exchange is determined by multiplying together the factors of $-1$ shown in the diagram with an additional factor of $-1$ from the crossing of fermion worldlines. }
\end{figure}

In the paired fermion model, a state on the composite site can only change through hops or exchanges of two fermions between neighboring sites. For example when one of the four states in (\ref{4states}) on a composite site moves to its neighboring empty site, both of its fermionic components must hop together. Since no fermion worldlines cross during such bosonic hops, there is no negative sign introduced during the hop. In contrast, when neighboring states exchange two fermions between them, there is usually a sign change since fermion worldlines cross each other. However, due to the definition of the bosonic state this is not always the case. One can work out the sign change during the exchange pictorially as we illustrate by considering six different examples of exchanges in Fig.~\ref{fig6}.  The complete list of fermion exchange processes involving two boson exchanges and the corresponding sign of the matrix element is given in table \ref{tab3}.  In the last row we also show the rules through which paired fermions on the composite chain interact with unpaired fermions on the fundamental chain. Note that the $s$ type particles do not interact with the fermion on the fundamental site, but the $t$ type particles do. Finally, we note that $H_\mu$ annihilates the $t$-type particles, but not the $s$-type particles. Since $H_\mu^2$ acts like a chemical potential for the $s$-type particles, it does not introduce negative signs in the worldline representation.

\begin{table}[h]
\begin{tabular}{cc||cc}
\hline
Process & Sign & Process & Sign \\
$t^+ \ t^-\ \longleftrightarrow s\ s$ & - &
$t^+ \ t^-\ \longleftrightarrow t^0\ t^0$ & + \\
$t^+ \ t^0\ \longleftrightarrow t^0\ t^+$ & - &
$t^+ \ s\ \longleftrightarrow s\ t^+$ & - \\
$t^- \ t^0\ \longleftrightarrow t^0\ t^-$ & - &
$t^- \ s\ \longleftrightarrow s\ t^-$ & - \\
$t^0 \ t^0\ \longleftrightarrow s\ s$ & - &
$t^0\  s\ \longleftrightarrow s\ t^0$ & - \\ \hline
$t^+ \ \da \ \longleftrightarrow t^0 \ \ua$ & - &
$t^- \ \ua \ \longleftrightarrow t^0\ \da$ & + \\
$\ua \ \ \da \ \longleftrightarrow \da \ \ \ua$ & - \\
\hline
\end{tabular}
\caption{\label{tab3} Signs of transfer matrix elements associated with off diagonal processes in the worldline representation. While every worldline crossing produces a negative sign, pair creation and annihilation of $t$-type particles are also negative. }
\end{table}

\begin{figure}[h]
\includegraphics[width=0.6\textwidth]{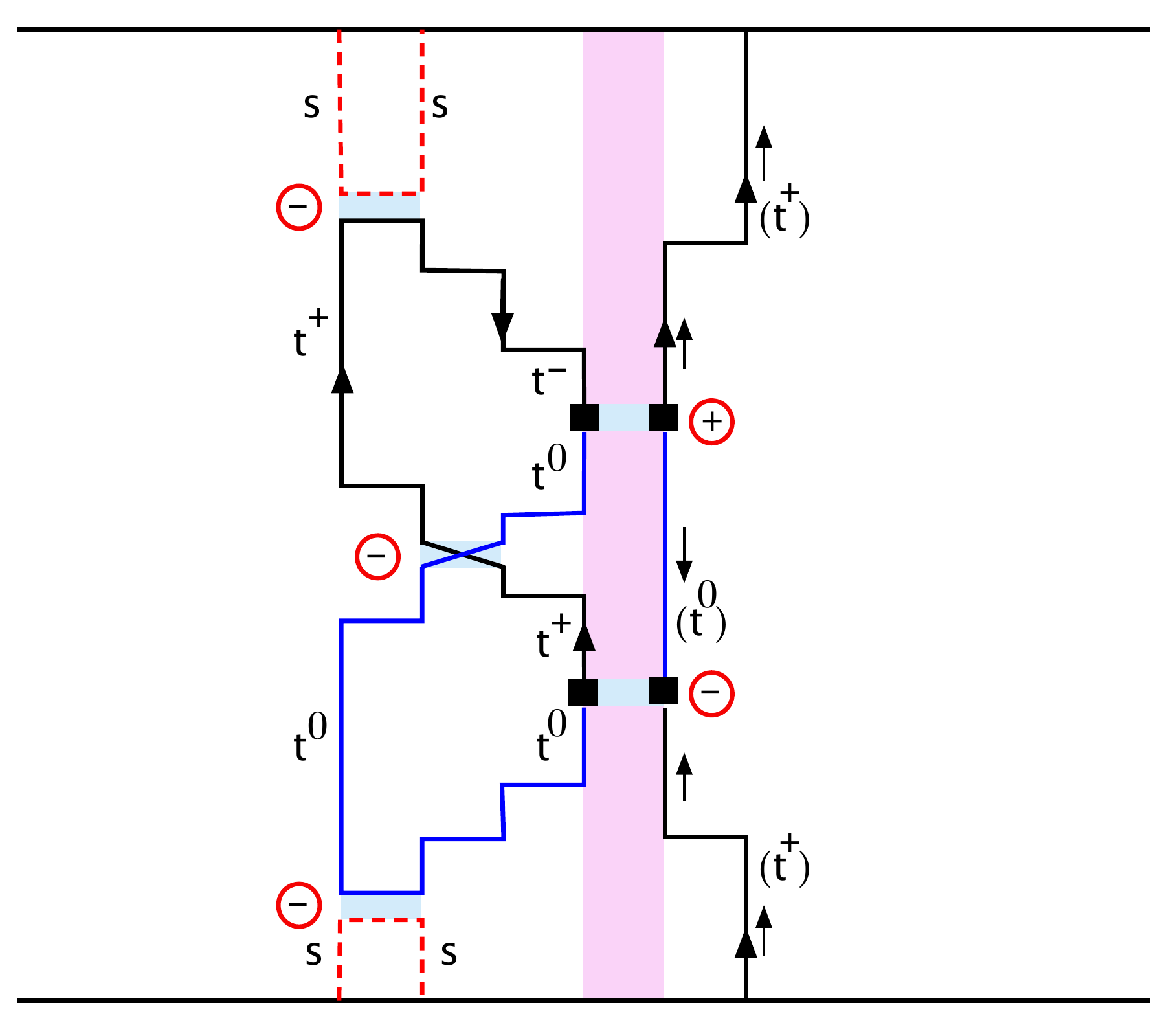}
\caption{\label{fig7} An illustration of a world line configuration in the paired fermion model with a single crossing of $s$ and $t$ type particles. As explained in the text, such configurations are still positive because we can map the $\uparrow$ spin on the fundamental sites with a $t^+$ particle and $\downarrow$ with a $t^0$ particle. With this mapping there are always an even number of crossings and even number of pair creation events with negative signs.}
\end{figure}

The sign of a worldline configuration can be computed using the information in table \ref{tab3}. But in order to define worldlines in the presence of interactions between paired fermions in the composite chain and unpaired fermions in the fundamental chain, we map a spin-up $\uparrow$ (spin-down $\downarrow$) fermion on the fundamental chain into a $t^+$ ($t^0$) particle. In other words, while pictorial drawing worldline configurations by following each of the four types of particles $t^+,t^-,t^0$ and $s$, only $t^+$ and $t^0$ particles travel to the fundamental chain from the composite chain, by transforming into $\uparrow$ and $\downarrow$ fermions respectively. We also view $t^+$ and $t^-$ as particles and anti-particles, and represent their worldlines as directed lines with an arrow pointing forward in time for $t^+$ and backwards in time for $t^-$. In contrast $t^0$ and $s$ are indicated through undirected worldlines. Using this pictorial representation every worldline configuration contains three types of closed loops, the $s$-loops, $t^0$ loops and the directed $t^+$-$t^-$ loops. Particles of each type cannot cross themselves, but can cross other types of particles. Note that in our model we do not allow $t^+$ and $t^-$ to cross each other since we forbid sites with four fermions. Thus every loop can only cross another loop of a different type. But, every crossing of worldlines produces a negative sign as seen from table \ref{tab3}. Since these lines are drawn on a two dimensional world sheet, such crossings always occur in pairs. The one-dimensional nature of the problem is important here. In addition to particle crossings, particles can be created or annihilated in pairs. Such events occur when world lines turn around in time. Interestingly we can associate negative signs with pair creation/annihilation of $t^0-t^0$ and $t^+-t^-$ particles, while keeping similar events for $s-s$ particles positive. These are also consistent with the rules in \ref{tab3}. Since pair-creation and annihilation events also come in pairs in every loop, they too cancel. Thus, all world line configurations are positive in the paired fermion model and the sign problem is absent.  An illustration of a world line configuration in the paired fermion is shown in Fig.~\ref{fig7}.

\section{The Quantum Spin Limit}
\label{sec5}

In the paired fermion model we allowed composite sites and fundamental sites to be empty. If we eliminate all empty sites we recover our original frustrated quantum spin model. This means we could have chosen an appropriate local basis in our original spin model to avoid all sign problems. In this section we construct this basis directly. Taking clues from the paired fermion model, we combine the chains $a=1,2$ into a single chain with a four dimensional Hilbert space. We then choose the eigenstates of the total spin operator $\bfS_i = \bfS_{1,i}+\bfS_{2,i}$ as the complete basis to expand the partition function. The lattice structure of the model is the same as Fig.~\ref{fig3}, where the fundamental sites are denoted by circles and contain spin-half states, while the composite sites are denoted by squares and contain the $s$ and the three $t$ states.

Given two independent quantum spin-half operators $\bfS_1$ and $\bfS_2$, we can label the four-dimensional Hilbert space on a composite site either through the eigenstates of $S_1^z$ and $S_2^z$ as $\left|\uparrow\uparrow\right\rangle$, $\left|\uparrow\downarrow\right\rangle$, $\left|\downarrow\uparrow\right\rangle$, $\left|\downarrow\downarrow\right\rangle$, or through the eigenstates of the total spin $\bfS_i =\bfS_{1,i}+\bfS_{2,i}$ states as the spin singlet $\left|s\right\rangle$ and the three spin triplets $|t^0\rangle, \ |t^+\rangle,\ |t^-\rangle$. Using the insight from the paired fermion model, we define these states through the expressions
\begin{equation}
|s\rangle = \frac{1}{\sqrt{2}} (\left|\uparrow\downarrow\right\rangle - \left|\downarrow\uparrow\right\rangle), \ \ 
|t^0\rangle = \frac{1}{\sqrt{2}} (\left|\uparrow\downarrow\right\rangle + \left|\downarrow\uparrow\right\rangle), \ \ 
|t^+\rangle = \left|\uparrow \uparrow\right\rangle,\ \ |t^-\rangle = -\left|\downarrow \downarrow\right\rangle.\ \ 
\label{sbasis}
\end{equation}

The extra negative sign in the definition of $|t^-\rangle$ is the remnant of the negative sign introduced in (\ref{basisc}) and will be useful in mapping the signs in the quantum spin model to the signs in the paired fermion model.

Let us now construct all the matrix elements of the Hamilton operator (\ref{smodel}) in the above basis. We divide the operator into three parts $H = H_L+H_R+H_I$ for convenience, such that 
\begin{eqnarray}
H_L \ &=& \  \lambda_s \sum_{a=1,2} \sum_{i=0}^{N-1} \bfS_{a,i} \cdot \bfS_{a,i+1},\ \nonumber \\
H_R \ &=&\  \lambda_s \sum_{i=0}^{N-1} \bfS_{3,i} \cdot \bfS_{3,i+1},\ \nonumber \\
H_I\  &=&\  \alpha_s \big(\bfS_{1,0}\cdot \bfS_{2,0} + (\bfS_{1,0}+\bfS_{2,0})\cdot \bfS_{3,0}\big).
\label{shamop}
\end{eqnarray}
The action of each term of the Hamiltonian on the nearest neighbor states in the basis (\ref{sbasis}) is given in table \ref{tab4}. From this information it is easy to read off the matrix elements. We then collect all the diagonal terms in an operator defined as $H_D$ and all the off diagonal terms in $H_O$. We can then expand the partition function in the CT-INT formulation as before (see (\ref{ctint}))
\begin{equation}
Z = \sum_k \int_T dt_1 dt_2...dt_k \mathrm{Tr}\Bigg(\mathrm{e}^{-(\beta-t_1)H_D}(-H_O)\mathrm{e}^{-(t_1-t_2)H_D}(-H_O)...(-H_O)...(-H_O)\mathrm{e}^{-t_kH_\mu}\Bigg).
\end{equation}
If we expand the trace in the basis (\ref{sbasis}), the potential negative signs can only come from off diagonal terms. Since the chosen basis naturally defines a worldline configuration involving $s$ and $t$-type particles on composite sites and spins on fundamental sites, they are identical to the paired fermion model. Further, due to our choice of the basis, even the local negative signs from the off diagonal terms are exactly the same as those given in Table \ref{tab3}. Hence for the same reasons as already discussed in the paired fermion model, all configurations weights are positive and the sign problem is absent.

\begin{table}
\begin{tabular}{|c|c|}
\hline
$H_{L,ij}\ (|t^+\rangle \otimes |t^+\rangle) = \frac{\lambda_s}{2}\ (|t^+\rangle \otimes |t^+\rangle)\ \ \ $
&
$\ \ \ H_{L,ij}\ (|t^-\rangle \otimes |t^-\rangle) = \frac{\lambda_s}{2}\ (|t^-\rangle \otimes |t^-\rangle)$
\\
\hline
\multicolumn{2}{|c|}{$H_{L,ij}\ (|t^+\rangle \otimes |t^-\rangle) = -\frac{\lambda_s}{2}\ (|t^+\rangle \otimes |t^-\rangle) 
- \frac{\lambda_s}{2}\ [(|t^0\rangle \otimes |t^0\rangle) - (|s\rangle \otimes |s\rangle)]$} \\
\hline
\multicolumn{2}{|c|}{$H_{L,ij}\ (|t^-\rangle \otimes |t^+\rangle) = -\frac{\lambda_s}{2}\ (|t^-\rangle \otimes |t^+\rangle) - \frac{\lambda_s}{2}\ 
[(|t^0\rangle \otimes |t^0\rangle) - (|s\rangle \otimes |s\rangle)]$} \\
\hline
\multicolumn{2}{|c|}{$
H_{L,ij}\ (|t^0\rangle \otimes |t^0\rangle) = \frac{\lambda_s}{2} (|s\rangle \otimes |s\rangle) 
- \frac{\lambda_s}{2}\ [(|t^+\rangle \otimes |t^-\rangle) \ +\ (|t^-\rangle \otimes |t^+\rangle)]$} \\
\hline
\multicolumn{2}{|c|}{$
H_{L,ij}\ (|s\rangle \otimes |s\rangle) = \frac{\lambda_s}{2}\ (|t^0\rangle \otimes |t^0\rangle) + \frac{\lambda_s}{2}\ [(|t^+\rangle \otimes |t^-\rangle) \ +\ 
(|t^-\rangle \otimes |t^+\rangle)]$} \\
\hline
$H_{L,ij}\ (|t^+\rangle \otimes |t^0\rangle) = \frac{\lambda_s}{2}\ (|t^0\rangle \otimes |t^+\rangle)\ \ \ $ 
&
$\ \ \ H_{L,ij}\ (|t^+\rangle \otimes |s\rangle) = \frac{\lambda_s}{2}\ (|s\rangle \otimes |t^+\rangle)$\\
\hline
$H_{L,ij}\ (|t^-\rangle \otimes |t^0\rangle) = \frac{\lambda_s}{2}\ |t^0\rangle \otimes |t^-\rangle)\ \ \ $
&
$\ \ \ H_{L,ij}\ (|t^-\rangle \otimes |s\rangle) = \frac{\lambda_s}{2}\ (|s\rangle \otimes |t^-\rangle)$ \\
\hline
$H_{L,ij}\ (|t^0\rangle \otimes |t^+\rangle) = \frac{\lambda_s}{2}\ (|t^+\rangle \otimes |t^0\rangle)\ \ \ $
&
$\ \ \ H_{L,ij}\ (|t^0\rangle \otimes |t^-\rangle) = \frac{\lambda_s}{2}\ (|t^-\rangle \otimes |t^0\rangle)$ \\
\hline
$H_{L,ij}\ (|t^0\rangle \otimes |s\rangle) = \frac{\lambda_s}{2}\ (|s\rangle \otimes |t^0\rangle)\ \ \ $
&
$\ \ \ H_{L,ij}\ (|s\rangle \otimes |t^0\rangle) = \frac{\lambda_s}{2}\ (|t^0\rangle \otimes |s\rangle)$ \\
\hline
$H_{L,ij}\ (|s\rangle \otimes |t^+\rangle) = \frac{\lambda_s}{2}\ (|t^+\rangle \otimes |s\rangle)\ \ \ $
&
$\ \ \ H_{L,ij}\ (|s\rangle \otimes |t^-\rangle) = \frac{\lambda_s}{2}\ (|t^-\rangle \otimes |s\rangle)$
\\
\hline \hline
$H_I\ (|t^+\rangle \otimes \left|\uparrow\right\rangle) = \frac{3\alpha_s}{4} (|t^+\rangle \otimes \left|\uparrow\right\rangle)\ \ \ $
&
$\ \ \ H_I\ (|t^-\rangle \otimes \left|\downarrow\right\rangle) = \frac{3\alpha_s}{4} (|t^-\rangle \otimes \left|\downarrow\right\rangle)$ \\
\hline
$H_I\ (|s\rangle \otimes \left|\uparrow\right\rangle) = -\frac{3\alpha_s}{4} (|s\rangle \otimes \left|\uparrow\right\rangle)\ \ \ $
&
$\ \ \ H_I\ (|s\rangle \otimes \left|\downarrow\right\rangle) = -\frac{3\alpha_s}{4} (|s\rangle \otimes \left|\downarrow\right\rangle)$  \\
\hline
\multicolumn{2}{|c|}{$
H_I\ (|t^+\rangle \otimes \left|\downarrow\right\rangle) = -\frac{\alpha_s}{4} (|t^+\rangle \otimes \left|\downarrow\right\rangle) 
+ \frac{\alpha_s}{\sqrt{2}} |t^0\rangle \otimes \left|\uparrow\right\rangle$} 
\\
\hline
\multicolumn{2}{|c|}{$
H_I\ (|t^-\rangle \otimes \left|\uparrow\right\rangle) = -\frac{\alpha_s}{4} (|t^-\rangle \otimes \left|\uparrow\right\rangle) 
- \frac{\alpha_s}{\sqrt{2}} |t^0\rangle \otimes \left|\downarrow\right\rangle$}
\\
\hline
\multicolumn{2}{|c|}{$
H_I\ (|t^0\rangle \otimes \left|\uparrow\right\rangle) = \frac{\alpha_s}{4} (|t^0\rangle \otimes \left|\uparrow\right\rangle) + \frac{\alpha_s}{\sqrt{2}} |t^+\rangle \otimes \left|\downarrow\right\rangle$}
\\
\hline
\multicolumn{2}{|c|}{$H_I\ (|t^0\rangle \otimes \left|\downarrow\right\rangle) = \frac{\alpha_s}{4} (|t^0\rangle \otimes \left|\downarrow\right\rangle) - \frac{\alpha_s}{\sqrt{2}} |t^-\rangle \otimes \left|\uparrow\right\rangle$}
\\
\hline \hline
$H_{R,ij}\ (\left|\uparrow\right\rangle \otimes \left|\uparrow\right\rangle) = \frac{\lambda_s}{4} (\left|\uparrow\right\rangle \otimes \left|\uparrow\right\rangle)$ &
$H_{R,ij}\ (\left|\downarrow\right\rangle \otimes \left|\downarrow\right\rangle) = \frac{\lambda_s}{4} (\left|\downarrow\right\rangle \otimes \left|\downarrow\right\rangle)$ \\
\hline
\multicolumn{2}{|c|}{$
H_{R,ij}\ (\left|\uparrow\right\rangle \otimes \left|\downarrow\right\rangle) = -\frac{\lambda_s}{4} (\left|\uparrow\right\rangle \otimes \left|\downarrow\right\rangle) + \frac{\lambda_s}{2}\ (\left|\downarrow\right\rangle \otimes \left|\uparrow\right\rangle)$}
\\ 
\hline
\multicolumn{2}{|c|}{$
H_{R,ij}\ (\left|\downarrow\right\rangle \otimes \left|\uparrow\right\rangle) = -\frac{\lambda_s}{4} (\left|\downarrow\right\rangle \otimes \left|\uparrow\right\rangle) + \frac{\lambda_s}{2}\ (\left|\uparrow\right\rangle \otimes \left|\downarrow\right\rangle)$} \\
\hline
\end{tabular}
\caption{\label{tab4} Action of the various terms in the quantum spin Hamiltonian (\ref{shamop}) on the basis states (\ref{sbasis}). }
\end{table}

\section{Conclusions}
\label{sec6}

In this work we solved the sign problem of a simple frustrated quantum spin model in which three quantum spin chains interact through a frustrated antiferromagnetic coupling. The solution required the use of a modified spin basis involving more than one lattice site of the original model. The idea behind the solution was obtained by starting from a fermion model that reproduced the spin model in the strong coupling limit. We then converted the fermion model into a non-homogeneous one dimensional problem. The solution emerged in the limit where fermions remain paired (into $s-$ and $t$-type bosons) on half of the space but remain unpaired in the other half. The frustrating interaction involved exchanging bosons and fermions at the impurity site. We showed that worldline configurations of bosons and fermions remain positive even in the presence of such an exchange. The solution remains valid even when the bosons and fermions are not present on every site as required in the original spin model. Thus, in a way we have found a solution to an extended class of problems that go beyond the original spin model. Another important lesson we learned is that although the frustration that produces the sign problem is localized, it seemed necessary to change the basis in a macroscopic region in order to solve the sign problem completely. All local reformulations of the basis we tried on a few sites close to the frustration did not yield a solution. This seems consistent with earlier findings \cite{PhysRevB.92.195126}.

Our solution should be extendable to other models as long as they can can be reduced to a series of one dimensional problems that may be coupled through interactions and that are diagonal in the worldline representation. It is also likely that other simple frustrated models can be solved using similar ideas. Unfortunately worldline configurations in our model with unpaired fermions can still have negative weights. The solution will need resummation of fermion world lines, which we did not attempt in this work. Perhaps ideas like the meron cluster solution \cite{PhysRevLett.83.3116} and the fermion bag approach  \cite{Chandrasekharan:2013rpa} would be helpful in this regard. It is also likely that solutions to fermionic models on ladder geometries could emerge from our work. Finally, it would also be interesting to explore if partition functions of more complex models in higher dimensions could be reduced to a sum of partition functions of simpler models of the type encountered here. 

\section*{Acknowledgments}

We would like to thank helpful conversations with Kedar Damle, Diptiman Sen and Uwe-Jens Wiese. SC would like to thank the Center for High Energy Physics at the Indian Institute of Science for hospitality, where part of this work was done. The material presented here is based upon work supported by the U.S. Department of Energy, Office of Science, Nuclear Physics program under Award Number DE-FG02-05ER41368. EH is also supported by a National Physical Science Consortium fellowship.

\bibliography{ref}

\end{document}